\documentclass[aip, reprint]{revtex4-1}

\usepackage[utf8]{inputenc}
\usepackage[T1]{fontenc}
\usepackage[english]{babel}

\usepackage{graphicx}
\usepackage{amsmath, amssymb}
\usepackage{mathptmx}
\usepackage[binary-units=true]{siunitx}
\usepackage{chemfig}
\usepackage[capitalise]{cleveref}
\usepackage{todonotes}
\begin{document}


\title{Alumina coating for dispersion management in ultra-high Q microresonators}

\author{Marvyn Inga}
\author{La\'{i}s Fujii}
\affiliation{Photonics Research Center, University of Campinas, Campinas 13083-859, SP, Brazil}
\affiliation{Applied Physics Department, Gleb Wataghin Physics Institute, University of Campinas, Campinas 13083-859, SP, Brazil}

\author{Jos\'{e} Maria C. da Silva Filho}
\affiliation{Applied Physics Department, Gleb Wataghin Physics Institute, University of Campinas, Campinas 13083-859, SP, Brazil}

\author{Jo\~{a}o Henrique Quintino Palhares}
\author{Andre Santarosa Ferlauto}
\affiliation{CECS, Federal University of ABC, Santo Andr\'{e} 09210-580, SP, Brazil}

\author{Francisco C. Marques}
\affiliation{Applied Physics Department, Gleb Wataghin Physics Institute, University of Campinas, Campinas 13083-859, SP, Brazil}

\author{Thiago P. Mayer Alegre}
\author{Gustavo Wiederhecker}
\email[]{gsw@unicamp.br}
\affiliation{Photonics Research Center, University of Campinas, Campinas 13083-859, SP, Brazil}
\affiliation{Applied Physics Department, Gleb Wataghin Physics Institute, University of Campinas, Campinas 13083-859, SP, Brazil}

\date{\today}

\begin{abstract}
Silica optical microspheres often exhibit ultra-high quality factors, yet, their  group velocity dispersion, which is crucial for nonlinear optics applications, can only be coarsely tuned. We experimentally demonstrate that group-velocity dispersion of a silica microsphere can be engineered by coating it with conformal nanometric layers of alumina, yet preserving its ultra-high optical quality factors (\num{\sim e7}) at telecom wavelengths. Using the atomic layer deposition technique for the dielectric coating, which ensures nm-level thickness control, we not only achieve a fine dispersion tailoring but also maintain a low surface roughness and material absorption to ensure a low optical loss. Numerical simulations supporting our experimental results show that the alumina layer thickness is a promising technique for precise tuning of group-velocity dispersion. As an application we demonstrate the generation of Kerr optical frequency combs, showing that the alumina coatings can also sustain the high optical intensities necessary for nonlinear optical phenomena.
\end{abstract}


\maketitle 

\section{Introduction}\label{Introduction}

Despite the consistent progress in mitigating optical losses within integrated photonics microcavities~\cite{Ji2017Ultra-low-lossThreshold, Wu2018}, fused-silica microspheres still rank amongst the highest optical quality factor microresonators ever fabricated~\cite{Benson2016}. These ultra-high $Q$-factors, reported above \num{e9} at visible and near-infrared~\cite{ Collot1993,Vernooy1998} and around to \num{e8} at telecom wavelengths~\cite{Spillane2002,Jager2011,Webb2016}, resulting from silica's low material absorption and atomic-level surface roughness achieved by the standard thermal-fusion fabrication process. Those properties, combined with a relatively small modal volume, allowed microspheres to host pioneer low-power nonlinear optical phenomena, such as Raman lasing~\cite{Spillane2002, Andrianov2020}, third-harmonic generation~\cite{Dominguez-Juarez2011, Farnesi2014, Chen-Jinnai2016}, Kerr optical frequency combs~\cite{Webb2016, Zhu2019} and optomechanical effects~\cite{Bahl2011}. The efficiency of many nonlinear processes also depends on momentum conservation (phase-matching), which requires control over the group-velocity dispersion (GVD) of resonator modes. While other types of microresonators offer more geometric degrees of freedom for GVD control, e.g., height or width~\cite{Levy2010}; sidewall angle~\cite{Yang2016,Fujii_Inga2020}; or circumference shape control\cite{Fujii2020,Lin2015,Grudinin2015},
the diameter is the only geometric degree of freedom in microspheres. Despite the outstanding possibilities explored so far\cite{Sayson2017}, changing the diameter of microspheres fabricated with glass fusion techniques has critical challenges, such as pre-tapering of an optical fiber or extremely fine control over an electrical arc discharge or \chemfig{CO_2}-laser fusion~\cite{Fan2006, Maker2012} process. A straightforward approach to overcome this lack of geometrical parameters is to use dielectric coatings, as experimentally demonstrated by Risti\`{c} et al~\cite{Ristic2014c} and numerically investigated in multi-layer coated spheres ~\cite{Jin2017, Wang2018}. Nevertheless, experiments with coated microspheres have been hindered by rather low optical $Q$-factors, which have not to exceed the \num{e5} range~\cite{Ke2019}.
 
 Here we demonstrate unprecedented control of GVD in silica microspheres coated with alumina (\chemfig{Al_2O_3}) while preserving ultra-high $Q$-factors ($\sim\num{e7}$). A smooth alumina layer is obtained using atomic layer deposition (ALD) with nanometer-level thickness control.  The optical power-handling of the coated microspheres was also preserved and  allowed the generation of broadband  (\SI{250}{\nm}-wide) optical frequency combs. Given the conformal characteristics of the ALD technique, the demonstrated approach could be readily extended to other high-Q resonator geometries and material platforms.

\begin{figure*}[htbp!]
\includegraphics{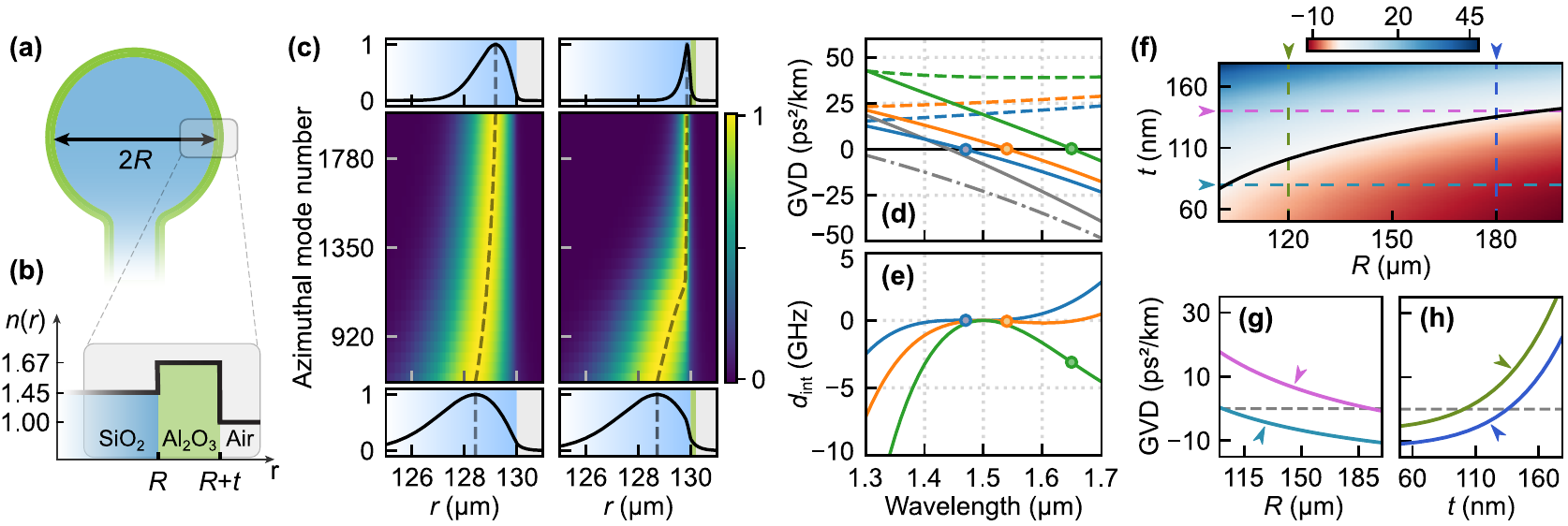}
\caption{\textbf{GVD control with alumina coating.} \textbf{(a)} Cross-section schematic of an alumina coated microsphere with radius $R$. \textbf{(b)} Surface refractive index step  due to a coating with  alumina layer with thickness $t$. \textbf{(c)} Radial distribution of the fundamental radial mode with increasing azimuthal mode number for a bare (left) and alumina coated (right) silica microsphere, the vertical axis corresponds to a frequency span from \SI{180}{\THz} ($m=\num{700}$) to \SI{490}{\THz} ($m=\num{2000}$). The grey dashed lines mark the peak position the optical field. \textbf{(d)} Comparison of numerically calculated total GVD (solid color lines) and purely waveguide GVD contribution (dashed color lines) for \SI{50}{\nm} (blue), \SI{110}{\nm} (orange) and \SI{150}{\nm} (green) thick alumina layers in a \SI{130}{\micro\meter} radius sphere. Gray curves are the material contribution of alumina (solid) and silica (dashed dot). \textbf{(e)} Residual dispersion curves corresponding to  the total GVD curves in (d). \textbf{(f)} Color-map representing the total GVD [\SI{}{ps
^2/km}] at $\lambda=$\SI{1550}{nm} as a function of microsphere radius ($R$) and alumina thickness ($t$) . The black line marks the zero GVD loci. \textbf{(g)} Curves of GVD versus $R$ for $t=\SI{80}{\nm}$ (cyan) and $t=\SI{140}{\nm}$ (purple) referenced in (f) with horizontal dashed lines. \textbf{(h)} Curves of GVD versus $t$ for $R=\SI{120}{\um}$ (olive) and $R=\SI{180}{\um}$ (blue) referenced in (f) with vertical dashed lines. All curves in this figure corresponds to TE polarization.}
\label{figure1}
\end{figure*}

\section{GVD control with alumina coating}
Group velocity dispersion in microresonators translates into a free-spectral range (FSR) that varies with frequency. The role of the alumina coating, illustrated in the schematic of \cref{figure1}(a,b), is to amplify the whispering gallery effect  as its higher refractive pulls the optical mode closer to the sphere surface. As contrasted in \cref{figure1}(c), the addition of a very thin alumina layer has a dramatic role on the modal profile. Such concentration towards the surface, not only increases the physical round-trip path but also slows down the optical mode, corresponding to an FSR that reduces at higher frequencies (higher azimuthal numbers) and, therefore, offers a higher degree of normal GVD. 
In order to contrast the radius and alumina thickness impact on the FSR, it is useful to explore the usual resonator optical frequency expansion~\cite{Herr2014b},
\begin{equation}
\nu(\mu) = \nu_0+d_1\mu+\frac{d_2}{2!}\mu^2+\frac{d_3}{3!}\mu^3+...,
\label{eq:omega_mu}
\end{equation}
where $\mu=m-m_0$ is the  azimuthal mode number relative to a reference mode ($m_0$); the coefficients $d_1$, $d_2$, $d_3$ represents, respectively, the FSR, the change of the FSR between successive modes, and so forth. To gain an insight on the geometry and material impact on the FSR dispersion, it is useful to relate the cavity dispersion coefficients $d_2$ of \cref{eq:omega_mu} to an equivalent GVD parameter $\beta_2$ [\si{s/m^2}] often used in the context of optical fibers and waveguides,
\begin{equation}
    d_2 = -\frac{c^3}{2\pi n_g^3 R^2}\beta_2,\\
    \label{eq:d2}
\end{equation}
where $R$ is the sphere radius and $n_g$ the optical mode group refractive index. Due to the overall $R^{-2}$ scaling of $d_2$, $\beta_2$ is more suitable quantity when evaluating the impact of radius or alumina thickness on the GVD. 

Despite the large index contrast of such microsphere resonators, further insight is possible if we break down to total GVD into independent  waveguide (geometric) and material dispersion contributions, $\beta_2\approx \beta_2^\text{(wg)}+\beta_2^\text{(mat)}$, an approximation commonly used in low index contrast waveguides\cite{Gloge1971}. The intuition of such decomposition and the impact of the alumina layer is confirmed in \cref{figure1}(d), where the total GVD of a microsphere is numerically calculated for three distinct alumina thicknesses (solid-color lines). Increasing from \SIrange{50}{150}{\nm}-thick alumina layer, the microsphere zero-dispersion wavelength can be tuned across $\approx$\SI{200}{\nm}. To confirm that the waveguide contribution (i.e.,  the surface confinement induced by the alumina's higher refractive index) is mostly responsible for the observed dispersion tuning, we show corresponding GVD curves where material dispersion of both silica ($n=\num{1.4446}$) and alumina ($n=\num{1.6171}$) were neglected. For reference, \cref{figure1}(d) also includes the material GVD of both alumina (solid-gray) and silica (dashed-dot-gray), where it is clear that the zero total GVD position is roughly given by the sum of silica's and waveguide contributions. As we discuss later, the role of alumina dispersion is negligible. In \cref{figure1}(e), we show the usual cavity residual dispersion curves, $d_\text{int}=\nu(\mu) - \nu_0-d_1\mu$, corresponding to the overall dispersion curves of \cref{figure1}(d), in these polynomial curves  the zero-GVD wavelength manifests as inflection points and positive (negative) concavity corresponds to anomalous (normal) dispersion. The  details on the numerical simulations are given in Supp. Info II.B.

It is clear thus that thin alumina layers can  lead to an efficient dispersion control, especially when a reduction of anomalous dispersion is desired, such as in broadband Kerr frequency comb generation~\cite{Saha2012,Li2017a}  or higher-order dispersion control~\cite{Sayson2019}. It remains elusive, however, whether this technique is indeed more effective than simply controlling the radius of microspheres. We investigate the interplay between these two dispersion control parameters in \cref{figure1}(f-h), where we focus on the transverse magnetic (TE) fundamental optical mode. The GVD colormap for TE-mode displayed in \cref{figure1}(f) shows that either parameters  are effective to access both normal and anomalous regimes. Actually, GVD varies more uniformly with radius than it varies with alumina thickness, e.g., below $t=\SI{80}{\nm}$ the GVD slope in \cref{figure1}(h) is rather flat. However, despite this apparent advantage of radius control, its imprecision level achieved with thermal the fusion technique is often on the order tens of \si{\um} range (See Suppl. Info. II.A.), while the degree of control over alumina thickness is readily available at \si{\nm}-level using an ALD process. This fine control knob is a key aspect to enable dispersion engineering when coating microresonators.

\section{High-Q alumina coated microspheres}

\begin{figure}[tbp!]
\includegraphics{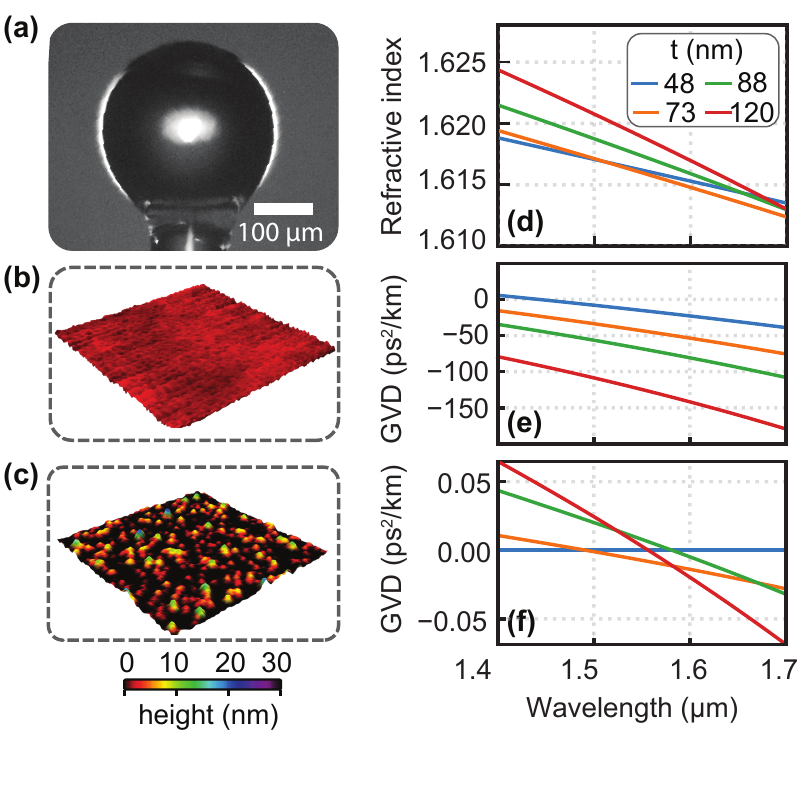}
\caption{\textbf{Optical properties and surface characterization of coated microspheres}. \textbf{(a)} Microscope image of a silica microsphere. Superficial characterization using AFM of \SI{3x3}{\um} sample areas reveal small RMS surface roughness for \textbf{(b)} bare silica microsphere $\sigma_\mathrm{SiO_2} = \SI{0.4}{\nm}$ and \textbf{(c)} \SI{30}{\nm}-thickness alumina coated microsphere $\sigma_\mathrm{Al_2O_3}=\SI{4}{\nm}$. \textbf{(d)} Refractive index of different alumina thickness obtained using the ellipsometry technique. \textbf{(e)} Material GVD contribution, $\beta_2
^\text{(mat)}$, obtained from (d). \textbf{(f)} Relative total GVD of a coated sphere ($R=$\SI{130}{\um}, $t=$\SI{60}{nm}) with the varying alumina refractive index properties (matching the curves in parts (d,f)). All curves were subtracted from the total GVD obtained with the material properties of the 48 nm curve (blue).}
\label{figure2}
\end{figure}
Despite recent demonstrations of coated microspheres for sensing applications~\cite{Mallik2017,Barucci2018a,Ke2019a} and GVD control~\cite{Ristic2014b}, their high optical loss -- associated with the sol-gel coating -- hinders their usage in the context of nonlinear optics. Here, we fabricated a series of silica microspheres with high-quality factor ($Q\sim\num{e7}$) coated with a conformal alumina layer.  They were fabricated using arc-fusion  fusion process of a standard optical fiber and coated using an atomic layer deposition technique. A typical microsphere obtained using this technique is shown in \cref{figure2}(a) (see Suppl. Info. II.A for details).
The alumina deposition using atomic layer deposition was performed in a BENEQ TFS 200 tool. The ALD technique ensures \si{\nm}-level thickness control and conformal covering of the spherical surfaces~\cite{Khanna2014} (see Suppl. Info. II.B for details). We directly characterized the microsphere surface quality using atomic force microscopy (AFM), as shown in \cref{figure2}(b,c).  From these representative roughness maps, it is possible to obtain the root-mean-square (RMS) roughness $\sigma_\text{\chemfig{SiO_2}}=\SI{0.4}{\nm}$ ($\sigma_\text{\chemfig{Al_2O_3}}=\SI{4}{\nm}$) and correlation length $B_\text{\chemfig{SiO_2}} = \SI{90.6}{\nm}$ ($B_\text{\chemfig{Al_2O_3}}=\SI{70.7}{\nm}$) for uncoated (alumina-coated) microspheres. Despite the rougher surfaces of the alumina-coated spheres, an analytical scattering-limited quality factor estimate based on a homogeneous sphere model predicts \cite{Vernooy1998} the roughness-limited $Q$-factors of \num{2e7} for the coated microsphere, in contrast with \num{3e9} for a pure silica microsphere with the same \SI{250}{\um} diameter at $\lambda=\SI{1550}{\nm}$. It is also worth noting that water adsorption often limit bare silica microsphere quality factors to the \num{e8} level \cite{Vernooy1998, Chen2017} (see Suppl. Info. II.B for details).

Another important aspect of nm-thick alumina layers is their  thickness-dependent refractive index~\cite{Shi2018}. Although this could hinder our dispersion management approach, we verify it to have minimal impact by coating plain silicon wafer pieces simultaneously with the each ALD deposition process. These coated pieces were used to perform optical ellipsometry and characterize the alumina layer thickness and refractive index. The refractive index frequency dependence is shown in \cref{figure2}(d) for four alumina layers with different thicknesses, \SIlist{48;73;88;120}{\nm}, as measured by ellipsometry (see Suppl. Info. II.B for details). The material refractive index Sellmeyer's fittings corresponding to each thickness are shown in \cref{figure2}(d). Indeed, there is a large variation of material GVD associated with each film, as shown in the solid colored curves of \cref{figure2}(e). Despite this substantial variation, the alumina material contribution only marginally affects the overall resonator GVD due to the small fraction of optical energy confined to the alumina layer (\SI{\sim 4}{\percent} according to numerical simulation). We confirm this negligible GVD impact by simulating an alumina coated microsphere with fixed  $R=$\SI{130}{\um} and $t=$\SI{60}{nm})  but varying the alumina refractive index dispersion according to the measured curves shown in \cref{figure2}(d); the total GVD curves for these resonators are so similar that they would visually overlap in wide vertical scale of \cref{figure2}(e). We highlight their small differences by subtracting from each total GVD curve, the one obtained using the refractive curve of the thinnest (\SI{48}{\nm}, blue curve) alumina; the variation is at most $\pm$\SI{0.05}{ps
^2/km}. Based on the dispersion slope of \cref{figure1}(d), $|\partial_\lambda\beta_2|\approx$\SI{0.1}{ps^2/km/nm}, the corresponding  zero-GVD wavelength shift is $\approx$\SI{0.5}{nm}, which confirms that alumina's GVD thickness dependence can indeed be neglected for the thin layers explored in this work. Yet, it should be accounted for if thicker layers (or shorter wavelenghts) are of interest.

\begin{figure}[tbp!]
\includegraphics{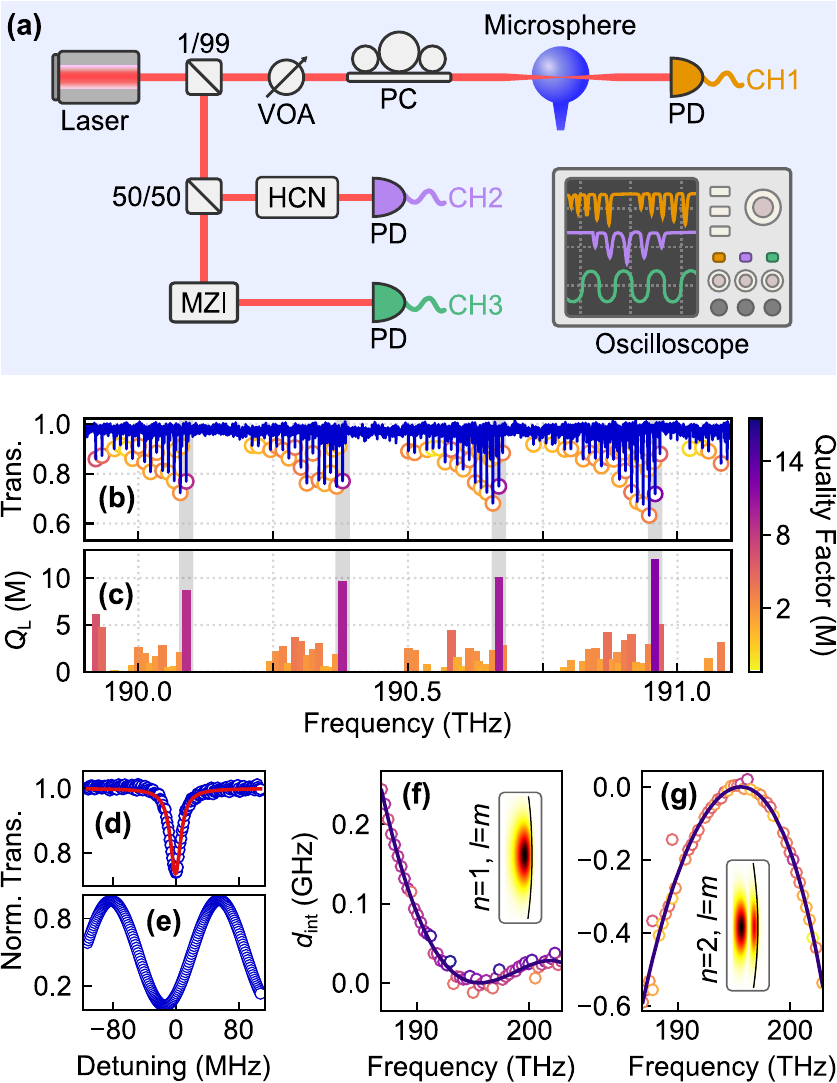}
\caption{\textbf{Optical characterization and dispersion measurements}. \textbf{(a)} A schematic of the experimental setup used for dispersion measurements. VOA: Variable attenuator, PC: Polarization controller, HCN: hydrogen cyanide gas cell, MZI: fiber Mach-Zhender interferometer, PD: Photodiode. \textbf{(b)} Transmission spectrum of \SI{73}{\nm} thick alumina coated microsphere and (c) Loaded $Q$-factor values corresponding to each resonance represented by color bars. \textbf{(d)} Lorentzian fit of the resonance with the highest loaded $Q$-factor value (\num{\sim 1.2e7}) in (b), centered at $\nu=\SI{190.96}{\THz}$. \textbf{(e)} Fiber-MZI transmission spectrum with $\text{FSR}=\SI{137\pm10}{\MHz}$. Applying polynomial fit to the  \textbf{(f)} fundamental mode ($n=1$, $l=m$) and \textbf{(g)} second radial order mode ($n=2$, $l=m$) yields $d_2$ equal to \SI{0.33\pm0.02}{MHz} and \SI{-1.4\pm0.1}{MHz} respectively. The insets in (f) and (g) show the transverse profile of the respective mode.}
\label{figure3}
\end{figure}

Optical transmission spectra of the microspheres were obtained by coupling light using a straight tapered fiber with \SI{\sim 3}{\micro\m} waist diameter in contact with the sphere surface near the equator, as schematically represented in \cref{figure3}(a). We tuned the optical resonances extinctions in the transmission spectra by adjusting the taper radius, which is known to significantly reduce the density of optical modes families appearing in transmission spectra~\cite{MohdNasir2016} (see Suppl. Info. I for details). Precise positioning and alignment were achieved using translation stages with \SI{50}{\nm} resolution (Suruga Seiki, KXC06020). In order to accurately measure the $Q$-factor and the relative frequency of the optical resonances, the transmission of a calibrated fiber-based (Corning SMF28) Mach–Zehnder interferometer (MZI $d_1=\SI{137\pm10}{\MHz}$, $d_2=\SI{0.5\pm0.1}{\Hz}$) and hydrogen cyanide (\chemfig{H^{13}C^{14}N}) gas cell were recorded along with the microsphere transmission spectra using an oscilloscope with memory length of 20.5~M points per channel (Keysight, 9245). The temporal resolution of the acquired traces enabled the resolution of $Q$-factors up to \num{e8} across the \SI{150}{\nm} tuning range of our external cavity diode laser (Tunics T100R, \SI{200}{\kHz} linewidth). The calibrated fiber-MZI based measurement was shown  to be accurate and reliable~\cite{Fujii_Inga2020, Fujii2020,DelHaye2009}.

An excerpt of the transmission spectrum for a \SI{73}{\nm} thick alumina-coated microsphere is shown in \cref{figure3}(b). Consecutive optical mode groups separated by a free spectral range ($\text{FSR}\simeq\SI{290}{\GHz}$) are excited. Each mode is identified by three integer numbers; the radial order ($n$), polar number ($l$) and azimuthal number ($m$). Due to the equatorial position of the tapered fiber, only modes concentrated around the equator ($|m|\lesssim l$) have are visible in the trasmission spectrum. \cref{figure3}(c) shows the loaded $Q$-factor for selected high extinction resonances. A typical \num{1.2e7} loaded $Q$-factor mode centered at \SI{190.96}{\THz} is shown in \cref{figure3}(d) along with the reference \SI{137}{\MHz} fiber-MZI trace in \cref{figure3}(e).

In order to extract the cavity GVD parameter ($d_2$) from the frequency-calibrated transmission spectra, we fitted \cref{eq:omega_mu}  (up to $d_3$) to a frequency versus mode number dataset~\cite{Fujii_Inga2020}. By removing the  offset ($\nu_0$) and subtracting the linear term ($d_1$) we obtain the experimental residual dispersion $d_\text{int}$, in which only the GVD ($d_2$) and higher-order terms are relevant. This detailed GVD characterization, combined with accurate 1-dimensional (1D) finite element simulations numerical simulations, allow us to clearly identify distinct radial orders, as shown in \cref{figure3}(f,g) for the $n=1$ and $n=2$ radial orders. While the polar order ($l-m$) are also discernible in such residual dispersion analysis, their dispersion profile are almost degenerate and are more easily resolved from the expected excentricity splitting easily identified in the transmission spectrum \cite{Righini2011}. We emphasize that this technique allows us to be certain about the identification of the $n=1$ radial order, otherwise, microsphere modes could be misidentified.

The experimental demonstration of GVD tailoring using different alumina-coating thicknesses are shown in \cref{figure4}(a), where curves corresponding to the measured residual dispersion ($d_\text{int}$) for alumina thickness ranging from \SIrange{48}{120}{\nm} are shown; for reference, the residual dispersion of a bare silica microsphere is also included. While both polarizations (TE and TM) experience a consistent reduction of the $d_2$ value (going from an anomalous ($d_2>0$) to a normal ($d_2<0$) dispersion), TE modes are more sensitive; we actually explored this feature to experimentally distinguish between the two polarization states.  Interestingly, near-zero GVD can be obtained, e.g. \SI{73}{\nm} (\SI{88}{\nm}) for TE (TM) polarization, confirming that ALD alumina can provide the fine control necessary for GVD control in high-$Q$ microresonators. In these regimes, the influence of the higher-order dispersion parameter becomes progressively more important, and an almost pure third-order dispersion ($d_3$) contribution can be obtained. Such a degree of control could be readily used for higher-order dispersion engineering~\cite{Parra-Rivas2014,Sayson2017}. The roughness impact of the alumina layer could also be noticed at thicker coatings. In \cref{figure4}(b) the average loaded quality factor for the fundamental mode across the measured range slightly deteriorates, reaching around \num{5e6} for the \SI{120}{\nm} layer. Although this trend is consistent with the previous reports of rougher surfaces in thicker alumina films~\cite{Weckman2015}, we believe that further optimization of the deposition process could lead to smoother coatings with potentially lower loss. 

\section{Optical Frequency comb generation}

\begin{figure*}[tbp!]
\includegraphics{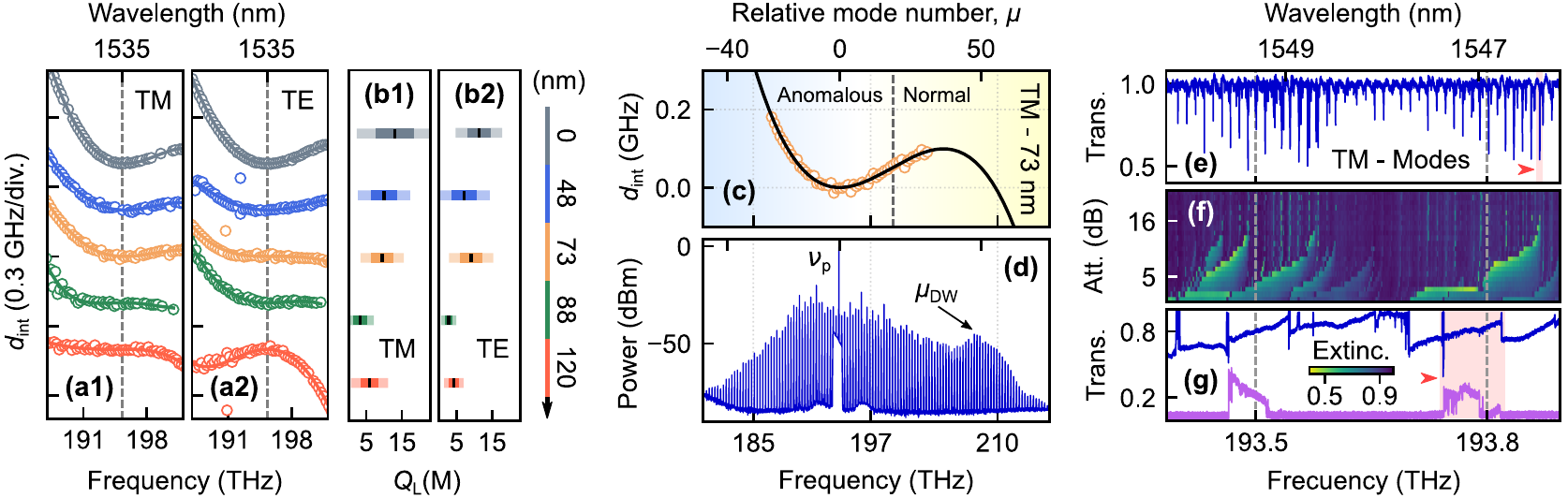}
\caption{\textbf{Dispersion control and frequency comb generation}. \textbf{(a)} Residual dispersion dependence on the alumina coating thickness for \textbf{(a1)} TM and \textbf{(a2)} TE polarization. Each curve was offset in the vertical axis by \SI{0.3}{\GHz}. \textbf{(b)} Mean loaded $Q$-factors (black marks)  for \textbf{(b1)} TM and \textbf{(b2)} TE polarization measured in the optical transmission window. The darkness (lightness) horizontal bar represent \SI{68}{\percent} (\SI{95}{\percent}) of the values lie within one (two) standard deviation. \textbf{(c)} Fitted TM-polarization dispersion curve for \SI{73}{\nm} thick alumina coated microsphere. Dashed line indicate the zero dispersion frequency (\SI{\sim 199.3}{\THz}). \textbf{(d)} Optical frequency comb with dispersive-wave signature around \SI{207,6}{\THz} produced in the \SI{73}{\nm} thick alumina coated microsphere. The pump frequency and power are $\nu_\text{p}=\SI{193,76}{\THz}$ (\SI{1547.2}{\nm}) and $P_\text{p}=\SI{65}{\mW}$, respectively. \textbf{(e)} Linear transmission spectrum and identification of the mode that produced the OFC in (d). This shaded mode appeared at \SI{193.87}{\THz} and had a $Q$-factor of \num{1.7e7}. \textbf{(f)} The colormap represent the transmission spectrum for different attenuation values used for mode identification. \textbf{(g)} High power (\SI{120}{mW}) bistable transmission spectrum before attenuation (in blue). The comb power, in purple, show exactly where the OFCs had been generating.}
\label{figure4}
\end{figure*}

One important application of GVD engineering within microspheres is the generation of Kerr optical frequency combs. A key aspect when engineering dispersion for broadband comb generation is to control the ratio of second-order ($d_2$) to third-order dispersion ($d_3$). Indeed, cascaded four-wave mixing can phase match at modes distant from the pump wave giving rise to a dispersive wave (DW) emission, the azimuthal mode number satisfying this criterion is $\mu_\text{DW} = -3d_2/d_3$, also known as  Cherenkov radiation in the context of mode-lock soliton Kerr combs~\cite{Brasch2016}. \cref{figure4}(c,d)  shows the residual dispersion curve fitted around the pump frequency ($\nu_\text{p}$) of an optical frequency (non-solitonic) comb generated in the \SI{73}{\nm} thick alumina-coated microsphere (\SI{65}{mW} CW-pump power). 
An increase in the comb intensity near \SI{207.6}{\THz} agrees well with the residual dispersion zero-crossing  ($d_\text{int}=0$) shown in \cref{figure4}(c).  Indeed, around the pump-mode position ($\nu_\text{c}=\SI{193.87}{\THz}$) the dispersion paramteres are $d_2=\SI{440\pm10}{\kHz}$, $d_3 = \SI{-24\pm1}{\kHz}$. If we replace this dispersion parameters values for the spectral position of the Cherenkov radiation peak, we find $\mu_\text{DW} = \num{55}$.  These results suggest that alumina coating can indeed enable the control of higher-order GVD features, such as dispersive-wave emission. Finally, in \cref{figure4}(e,f,g) we show that it is possible to identify the optical pump mode generating the comb, despite the dense optical transmission spectrum of the microspheres. By attenuating the input pump power (\SI{120}{mW}) while simultaneously monitoring both pump transmission and comb power (using an optical waveshaper, FINISAR 4000A) we gradually move from a strongly bistable optical spectrum to GVD-traceable Lorentzian one, confirming the fundamental TM mode as the seed of the observed Kerr microcomb.


\section{Conclusions}

In summary, we experimentally demonstrated ultra-high quality factors in alumina-coated silica microspheres. Using alumina coatings with roughly 100~nm we could achieve a large degree of GVD control, significantly reducing the degree of anomalous GVD yet preserving their high $Q$-factors. Furthermore, the alumina coated films could sustain the higher optical intensities necessary for the generation of broadband optical frequency combs, which brings a new degree of freedom for dispersion control in the context of nonlinear photonics. A route for tuning of dispersive-wave emission or engineering higher-order dispersion comb generation is clearly opened~\cite{Sayson2017, Sayson2019}. Although a single coating layer does not readily generate blue and red-shifted dispersive waves, the presented technique could be extended with alternating dielectric layers to achieve a larger range of GVD control
\cite{Liang:16}.

\section*{Supplementary Material}
See \href{http://supplementary//material}{supplementary material} for simulation details and some experimental considerations. In summary, the 1D Maxwell eigenproblem solver is discussed and validated. Details about the fabrication and characterization of the silica microspheres, alumina coating are characterization are also discussed. Finally, a detailed discussion of the  fiber-MZI calibration and direct comparision between simulated and measured residual dispersion curves are presented.

\begin{acknowledgments}
We would like to acknowledge Dr. Yovanny Valenzuela for his support with numerical simulations, Dr. Sandro Marcelo Rossi for  his help in characterizing the MZI fiber dispersion and the Center for Research and Development in Telecommunications (CPqD) for providing access to their infrastructure. We also acknowledge the MultiUser Laboratory (LAMULT) at IFGW for support in the AFM measurements and the LCPNano at the Federal University of Minas Gerais (UFMG) for the support in the spectroscopic ellipsometer.
\end{acknowledgments}

\section*{Data Availability}
The data that support the findings of this study are openly available in ZENODO at http://doi.org/10.5281/zenodo.3932243, reference number \cite{Inga_zenodo_2020} (upon publication).

\section*{Funding}
This work was supported by São Paulo Research Foundation (FAPESP) through grants 2012/17610-3, 2012/17765-7, 2018/15580-6, 2018/15577-5, 2018/25339-4, 2018/21311-8; Coordenação de Aperfeiçoamento de Pessoal de Nível Superior - Brasil (CAPES) - Finance Code 001; National Council for Scientific and Technological Development (CNPq) grants 306297/2017-5, 435260/2018-9; and National Institute of Surface Engineering (INES/CNPq) grant 465423/2014-0. A.S.F. is a CNPq fellow.

\section*{References}
\bibliography{references.bib}

\end{document}


\lstset{basicstyle = \ttfamily, columns=fullflexible}
\title{%
  \large Supplementary material: Alumina coating for dispersion management in ultra-high Q microresonators \\
  \bigskip
  \bigskip
    }

\author{Marvyn Inga}
\author{La\'{i}s Fujii}
\affiliation{Photonics Research Center, University of Campinas, Campinas 13083-859, SP, Brazil}
\affiliation{Applied Physics Department, Gleb Wataghin Physics Institute, University of Campinas, Campinas 13083-859, SP, Brazil}

\author{Jos\'{e} Maria C. da Silva Filho}
\affiliation{Applied Physics Department, Gleb Wataghin Physics Institute, University of Campinas, Campinas 13083-859, SP, Brazil}

\author{Jo\~{a}o Henrique Quintino Palhares}
\author{Andre Santarosa Ferlauto}
\affiliation{CECS, Federal University of ABC, Santo Andr\'{e} 09210-580, SP, Brazil}

\author{Francisco C. Marques}
\affiliation{Applied Physics Department, Gleb Wataghin Physics Institute, University of Campinas, Campinas 13083-859, SP, Brazil}

\author{Thiago P. Mayer Alegre}
\author{Gustavo Wiederhecker}
\email[]{gsw@unicamp.br}
\affiliation{Photonics Research Center, University of Campinas, Campinas 13083-859, SP, Brazil}
\affiliation{Applied Physics Department, Gleb Wataghin Physics Institute, University of Campinas, Campinas 13083-859, SP, Brazil}

\date{\today}

\maketitle
\tableofcontents

\section{Optical modes in microspheres}

\begin{figure*}[htbp!]
\includegraphics{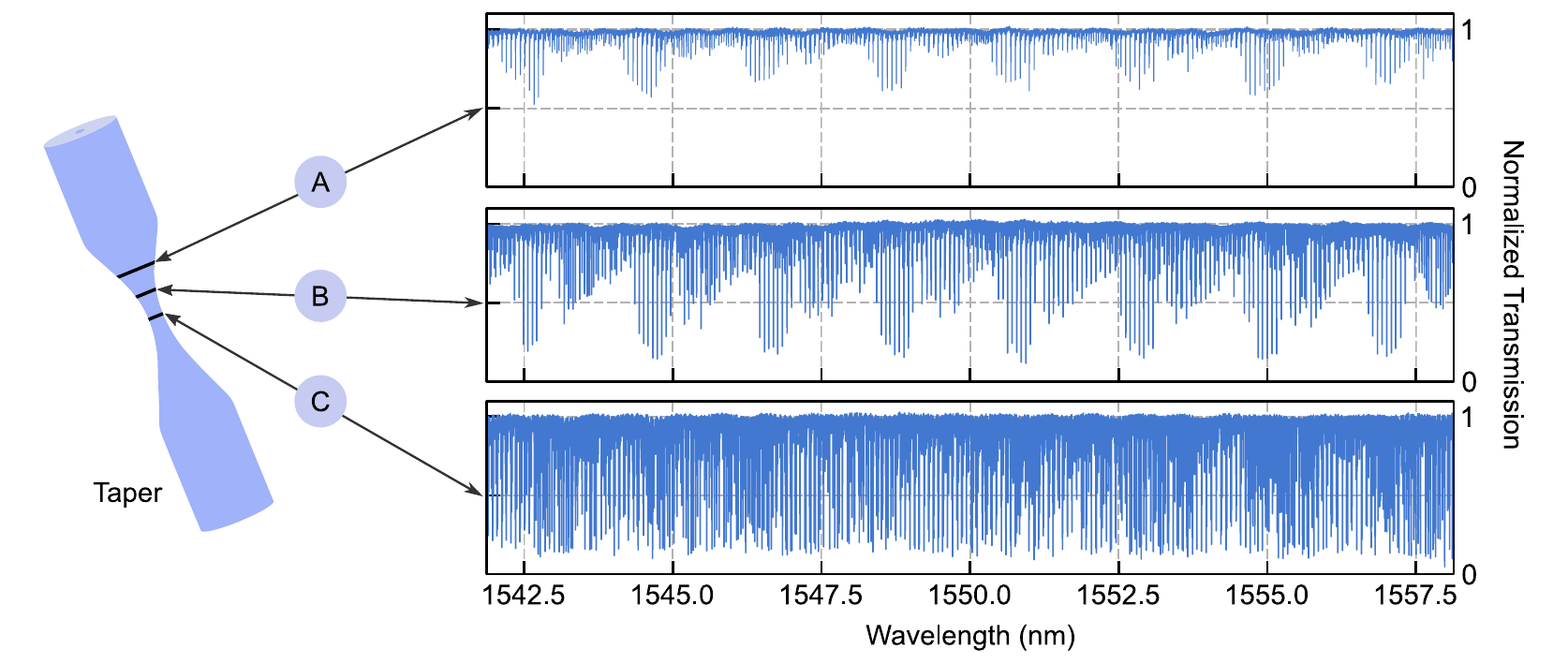}
\caption{Cleaning spectrum by fine-tuning of the taper diameter. Spectra A and B can be used to analyze dispersion and identify the first radial modes.}
\label{fig:3-Trans.Spectrum-vs-Diameter}
\end{figure*}

In a perfectly spherical resonator, there would be a large number of degenerate optical modes resulting from its rotational symmetry, rendering their distinction important for GVD characterization. Similarly to hydrogen atom, three numbers completely characterize an arbitrary optical mode of a sphere, $n,l,m$, respectively corresponding to radial, polar, and azimuthal orders; in addition to the two possible polarization states, Transverse Electric (TE) and Transverse Magnetic (TM). For each polar order $l$ there are $2l+1$  possible values for the azimuthal number $m$. Realistic microspheres, however, are eccentric and such degeneracy is lifted and readily observed in experiments \cite{Righini2011,Spillane2002}, furthermore, when light is coupled using tapered optical fibers, only modes localized along the equator ($m\lesssim l$) are efficiently excited. The precise positioning of the taper can thus be used to enhance the excitation of $l=m$ optical modes \cite{Knight:95}, In \cref{fig:3-Trans.Spectrum-vs-Diameter} we show an example of an optical spectrum for different taper radius showing the impact of phase-matching in the measured transmission spectrum.

\subsection{Solving 1-D Maxwell eigenproblem with FEM}
To numerically calculate GVD in microspheres, Maxwell's equations with proper boundary conditions must be sought. Following the separation of variables described by Jonhson et al \cite{Johnson1993}, the solution of the vector Helmholtz optical wave equation for a harmonic field at an angular frequency $\omega$ is reduced to solving the following ordinary differential equations for the radial Debye potentials,
\begin{subequations}
\begin{align}
    \frac{\mathrm{d}^{2} S_{l}(r)}{\mathrm{d} r^{2}}+\left[\frac{\omega^2 }{c^2} \epsilon(r,\omega)-\frac{l(l+1)}{r^{2}}\right] S_{l}(r) & =0\label{eq:TE}, \\
    \frac{\mathrm{d} }{\mathrm{d}r}\left(\frac{1}{\epsilon(r,\omega)}\frac{\mathrm{d} T_{l}(r)}{\mathrm{d}r}\right)+\left[\frac{\omega^2 }{c^2} -\frac{l(l+1)}{\epsilon(r,\omega)r^{2}}\right] T_{l}(r) & = 0\label{eq:TM},
\end{align}
\label{eq:debye}
\end{subequations}
\noindent where $S_l$ ($T_l$) is the radial Debye potential for the TE (TM) optical mode for a given polar number $l$, $\epsilon(r,\omega)$ is the frequency-dependent relative permittivity, and $c$ is the speed of light.  To satisfy Maxwell's boundary conditions of continuous tangential fields at a dielectric interface,  $S_l(r)$ and $\epsilon^{-1} T_{l}'(r)$ must be continuous. Note that TE (TM) polarization in a spherical resonator has a dominant electric-field along the polar (radial) spherical coordinate.


Although analytical solutions to the \cref{eq:TE,eq:TM} can be obtained for any combinations of piecewise continuous radial dielectric profiles, the characteristic equation for multi-layer dielectrics are not computationally cheap as they involve multiple evaluations of Bessel functions of very high order ($l\approx\num{600}$ for a \SI{100}{\um} at \SI{1500}{\nm}). 

Due to such limitations of the analytical approach, we use a 1-dimensional (1D) Finite Element Method (FEM) to solve \cref{eq:TE} or \cref{eq:TM} using COMSOL Multiphysics. Any of the COMSOL modules have these 1D equations already implemented. Therefore, we use the Galerkin method to derive the corresponding weak forms to be sought using the bare COMSOL Multiphysics package. Multiplying each of \cref{eq:TE,eq:TM} by a corresponding testing function $w_{l}(r)$ and integrating over the radial coordinate, the following residuals are obtained after integration by parts,
\begin{subequations}
\begin{align}
     \mathcal{R}_{\mathrm{TE}} &=\int dr\left[- \left(\frac{\mathrm{d} w_{l}(r)}{\mathrm{d}r} \right)\left(\frac{\mathrm{d} S_{l}(r)}{\mathrm{d}r} \right) + k^{2}  \epsilon(r,\omega)w_l(r) S_{l}(r) -l(l+1) \frac{w_l(r) S_{l}(r)}{r^{2}}\right] \label{eq:wTE}, \\
     \mathcal{R}_{\mathrm{TM}} &=\int dr\left[- \frac{1}{\epsilon(r,\omega)}\left(\frac{\mathrm{d} w_{l}(r)}{\mathrm{d}r} \right)\left(\frac{\mathrm{d} T_{l}(r)}{\mathrm{d}r} \right) + k^{2} w_l(r) T_{l}(r) -l(l+1) \frac{w_l(r) T_{l}(r)}{\epsilon(r,\omega)r^{2}}\right]\label{eq:wTM},
\end{align}
\label{eq:wdebye}
\end{subequations}
\noindent Note that the surface terms arising from integration by parts have been ignored, implying either Dirichlet or Neumann boundary conditions at the computational domain boundaries. In our implementation, since we are concerned with large spheres with negligible radiation losses, we do not implement any kind of absorbing boundary conditions. 

Once either of \cref{eq:wdebye} are discretized within COMSOL, they will be cast as a generalized matrix eigenvalue problem, and it must be chosen whether to treat it as a frequency eigenvalue problem ($\Lambda^2=\omega^2$ for a given $l$) or as a polar number eigenvalue problem ($\Lambda^2=l(l+1)$ for a given $\omega$). Although one might be tempted to use the more natural frequency eigenproblem, it becomes tricky to include the dielectric function frequency dispersion in this approach. COMSOL may handle this approach by linearizing the eigenvalue problem around a frequency of interest and self-consistently iterating the solution till convergence is obtained, however, it is not time-efficient as only one eigenvalue (mode) must be sought at a time and a few iterations might be necessary. Here we adopt the polar number eigenvalue approach, where a given frequency defines the frequency-dependent dielectric profile $\epsilon(r,\omega)$ and $l(l+1)$ is cast as the eigenvalue $\Lambda$, eliminating the need for iterations and allowing for solving multiple modes (radial orders) at once. The polar number is then recovered from the eigenvalue as $l=\frac{1}{2} \left(\sqrt{4 \Lambda
^2+1}-1\right)$, and a consequence of this choice is that the sought polar numbers are not integers. For the equatorial modes of interest, selected by phase-matching with the tapered fiber, the azimuthal numbers are given by $m=l$ and therefore we interpolate the solution at $l$-values. Using the latter approach, the values of $d_1,d_2,..,d_n$ values in the expansion (Eq. 1 in the main text) $$\nu(\mu) = \nu_0+d_1\mu+\frac{d_2}{2!}\mu^2+\frac{d_3}{3!}\mu^3+...$$ are recovered by fitting a polynomial function to the ($\omega/2\pi,l$) pairs with $l=m$.  

Example files implemented in COMSOL 5.5 are available in the ZENODO repository. The weak form contribution corresponding to \cref{eq:wdebye} are,
\begin{lstlisting}[breaklines]
     weakTE: -test(vr)*vr+k0^2*test(v)*v*material.def.epsilonr_iso-lambda^2*test(v)*v/r^2
    
    weakTM: -test(vr)*vr/material.def.epsilonr_iso+k0^2*test(v)*v-lambda^2*test(v)*v/r^2/material.def.epsilonr_iso
\end{lstlisting}
where we used the variable \lstinline{v} for the radial Debye potentials. Using the notation \lstinline{material.def.epsilonr_iso} for the dielectric constant $\epsilon(r)$ ensures that COMSOL will automatically account for domain-wise material properties; here the eigenvalue variable $\Lambda=$\lstinline{lambda} as internally defined by COMSOL.

\subsection{Connection between FSR and the GVD parameter}
Rigorously, the FSR of an optical resonator must be obtained directly from the characteristic equation derived from Maxwell's equation with the proper boundary conditions. However, one might interpret the cavity propagating modes (in contrast to standing-wave modes) as waveguide modes along a curvilinear coordinate. If we map this bent-waveguide mode  propagation direction into the sphere equator, $z\rightarrow R\varphi$, we can relate the waveguide propagation phase  with the cavity roundtrip phase, 
\begin{equation}
 \phi(\omega)= 2\pi R \beta(\omega), 
\end{equation}
where $\beta(\omega)$ is the waveguide propagation constant, which is commonly written in terms of the effective refractive index as $\beta(\omega)=(\omega/c)n_\text{eff}(\omega)$. To map the cavity FSR into the equivalent waveguide problem, we impose the cavity resonance condition,
\begin{equation}
    \phi(\omega_{m+1})-\phi(\omega_m)=2\pi
\end{equation}
Writing $\omega_{m+1}=\omega_m+\omega_\text{FSR}$, where $\omega_\text{FSR}$ corresponds exactly to one FSR, an expression for the FSR can be obtained up to first order, $$ \omega_\text{FSR}(\omega_m) = \frac{c}{R n_g(\omega_m)}=2\pi d_1,$$   
where $n_g(\omega)=n_\text{eff}+\omega \partial_\omega n_\text{eff}$ is the group refractive index. The variation of the FSR between two adjacent resonances can be obtained similarly by evaluating $\delta\omega_\text{FSR}= \omega_\text{FSR}(\omega_{m+1})-\omega_\text{FSR}(\omega_{m})$, which results 
\begin{equation}
 \delta\omega_\text{FSR}(\omega_m)=-\frac{c^3}{R^2}\frac{\beta_2(\omega_m)}{n_g
^3(\omega_m)}=2\pi d_2,
\label{eq:d2_analytic}
\end{equation}
where $\beta_2(\omega_m)\equiv\partial_{\omega}^2\beta(\omega)|_{\omega=\omega_m}=c^{-1}(2n_\text{eff}(\omega)+\omega\partial_{\omega}n_\text{eff})|_{\omega=\omega_m}.$ Similarly, higher-order variations of the FSR can be derived following the same approach. Yet, \cref{eq:d2_analytic} suffices to highlight the connection between the dispersion of the cavity FSR ($\delta\omega_\text{FSR}$) and the equivalent waveguide parameters and corresponds to Eq. (2) in the main text.

\begin{figure}
    \centering    \includegraphics{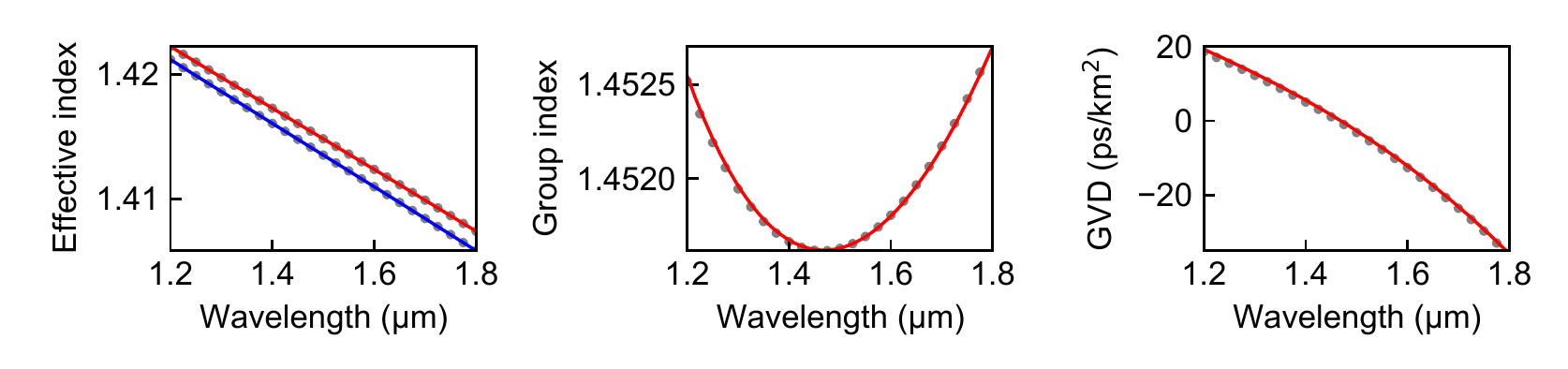}
    \caption{Numerical validation of the 1D FEM mode solver. (a) Comparison between analytical  (solid lines) and numerical (points) effective indexes for both TE (red) and TM (blue) modes, (b) group index for TE mode, and (c) GVD parameter for TE mode. TM mode group index and GVD curves are not shown due to an almost perfect match to the TE case.}
    \label{fig:analytical_validation}
\end{figure}

As a verification step of our 1D FEM solution and calculation of GVD parameters, we show in \cref{fig:analytical_validation} a comparison between relevant parameters calculated using the FEM solution (implemented in COMSOL) and analytical ones (implemented in Julia language), both are available in our the ZENODO repository.

\section{\label{section:experiment}Experimental considerations}
Here we give additional information about some experimental issues to supplement our results.

\subsection{Fused silica microspheres}

\begin{figure*}[htbp!]
\includegraphics{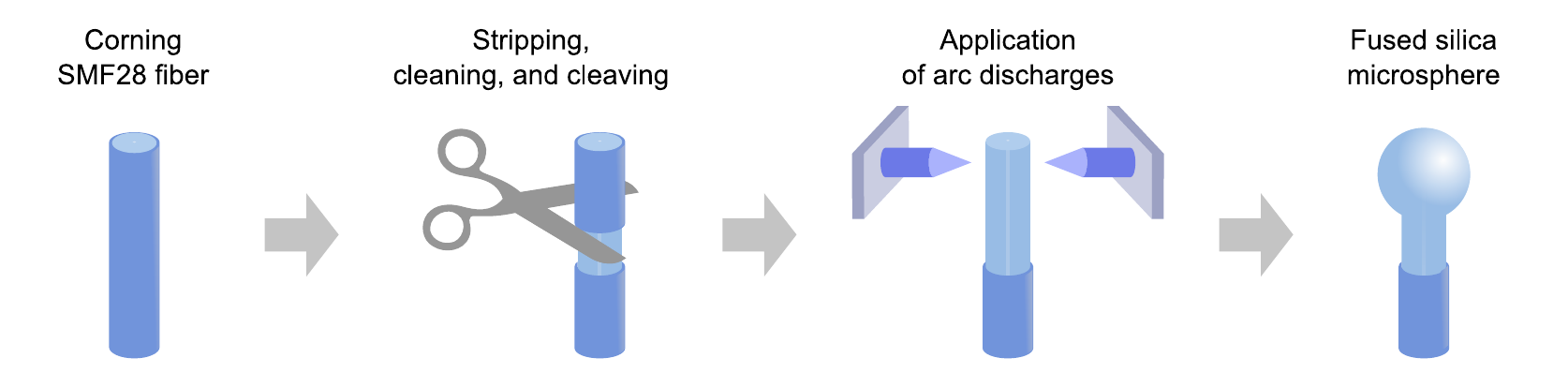}
\caption{Microspheres fabrication process.}
\label{fig:3-microsphere-fabrication}
\end{figure*}

Our fabrication procedure starts by cleaning and cleaving the tip of a standard telecom fiber (Corning SMF28) and subsequently melting it with a commercial fiber fusion splicer (Sumitomo Q101-CA) using electric arcs. As the heat is dissipated through the single-ended optical fiber taper, a sphere is formed due to molten silica's surface tension~\cite{Maker2012, Fan2006}. Control of the spherical shape and size is achieved by delivering several discharge arcs to the fiber tip, resulting in a small spherical dielectric cavity attached to a convenient optical fiber supporting stem, as illustrated \cref{fig:3-microsphere-fabrication}.
 
\begin{figure*}[htbp!]
\includegraphics{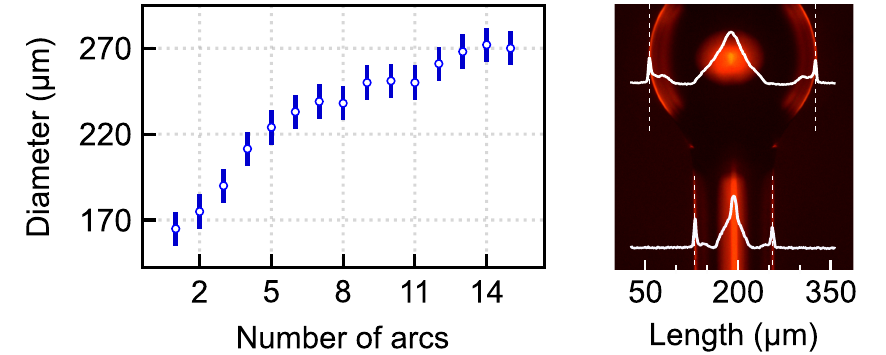}
\caption{Silica microsphere diameters versus the number of arcs. The photograph beside was calibrated to measure the diameter of a microsphere fabricated with 14 arcs.}
\label{fig:3-microsphere-diameter}
\end{figure*}

Silica microsphere diameters were measured by using a photograph taken from a frontal perspective. In each photo, we drew a line perpendicular to the microsphere stem and analyzed its plot profile using the software ImageJ. We used as a reference the external diameter of the Corning SMF28 fiber  (\SI{125}{\um}) to calibrated the dimensions of the photographs and thus found the microsphere diameters. \cref{fig:3-microsphere-diameter} shows the curve of the microsphere diameters versus the number of arcs applied. From here, it is clear to see that exists a saturation from 9 arcs onward. Another very relevant aspect is the difficulty of producing microspheres with a specific diameter,  often exits an interval of tens of \si{\um} around the desired diameter.
 
The eccentricity of the microspheres could be calculated from the equation given by Lai et al. \cite{Lai1990}:

\begin{equation}
    \epsilon = \frac{2l(l+1)}{2m+1}\left(\frac{\Delta\nu_\text{ecc}^{m, m+1}}{\nu_{nlm}^{0}}\right)
\end{equation}
where $\Delta\nu_\text{ecc}^{m,m+1} = |\nu_{nlm} - \nu_{n,l,m+1}|$ is the frequency splitting between successive azimuthal mode numbers, and can be obtained directly from the microsphere transmission spectrum. Considering the fundamental mode and typical homemade microsphere parameters: $R=\SI{130}{\um}$, $l=\num{716}$, $m=l$, $\Delta\nu_\text{ecc}^{m, m+1}\approx\SI{10}{\GHz}$, and $\nu_{nlm}^{0}\approx\SI{193}{\THz}$; we found an eccentricity of 3,7~\%. It can be translated too in a difference between the longest and shortest radio of \SI{4.8}{\um}.

\subsection{Atomic Layer Deposition and Sellmeyer parameters.}

\begin{figure*}[htbp!]
\includegraphics{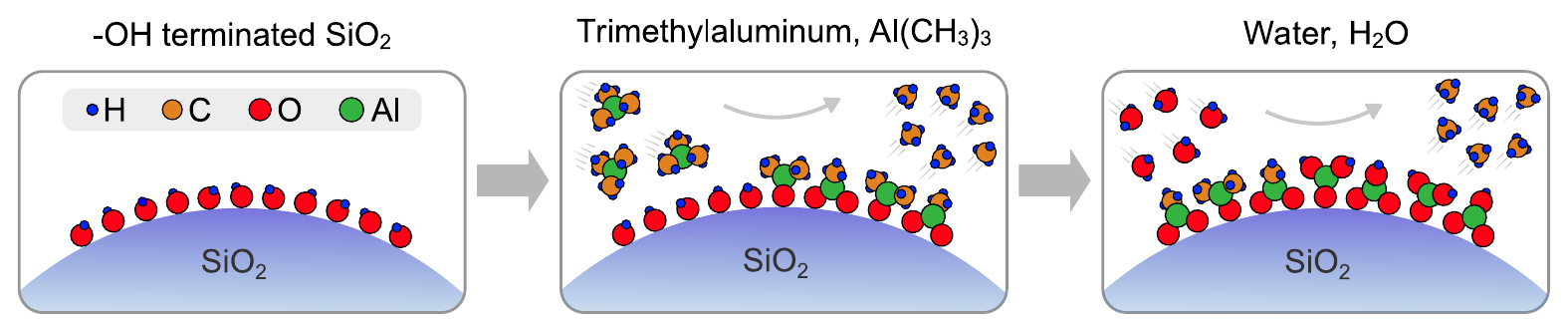}
\caption{ALD coating synthesis.}
\label{fig:3-ALD-process}
\end{figure*}

The cyclical process of the ALD is represented in \cref{fig:3-ALD-process}, and it is described by the next surface reactions \cite{Strempel2018},

\begin{align}
\ce{x||O-H + Al(CH_3)_3 (g)} & \ce{->} \ce{(||O-)_xAl(CH_3)_{3-x} + xCH4 (g)}\\
\ce{(||O-)_xAl(CH_3)_{3-x} + (3-x)H_2O} & \ce{->} \ce{(||O-)_xAl(OH)_{3-x} + (3-x)CH_4 (g)}
\end{align}
Taking $1\leq\ce{x}\leq 3$ we could understand all possible chemisorption mechanisms that TMA can undergo. The alumina formation was achieved by injecting two highly reactive vapor phase precursors into the ALD chamber. At \SI{250}{\celsius}, the first precursor (trimethylaluminium, TMA) reacted with the hydroxyl (-\chemfig{OH}) groups obtaining methane (\chemfig{CH_4}) as a product. Methane and excess of TMA were purged using an inert gas. The second precursor (water, \chemfig{H_2O}) produced a hydroxylated surface due to its disassociation. Products of this reaction were removed again starting thus a new cycle. Deposition parameters, tabulated in Table \ref{tab:FinalRecipe}, were optimized to minimize surface roughness of the deposited film.

\begin{table}[htbp!]
	\centering
	\caption{Parameters used in the alumina deposition.}
	\label{tab:FinalRecipe}
	\begin{ruledtabular}
	\begin{tabular}{lr}
		Parameter & Value \\ \hline
		Chamber temperature & \SI{250}{\celsius} \\
		TMA pulse time & \SI{0.15}{\s} \\
		Initial purge time & \SI{3}{\s} \\
		Water pulse time & \SI{0.15}{\s} \\
		Final purge time & \SI{5}{\s} \\
		Cycle duration & \SI{8.3}{\s} \\
		Growth rate per cycle   & $\sim\SI{0.8}{\angstrom}$ \\
	\end{tabular}
	\end{ruledtabular}
\end{table}

Spectroscopic ellipsometry measurements were performed using an M2000 Woollam with a rotating compensator in the range from \SIrange{246}{1689}{\nm} and using 4 incidence angles (55$^\circ$ to 70$^\circ$ in 5$^\circ$ steps).
The optical model used for the analysis consists of the Si substrate, native oxide layer (thickness of \SI{1.8}{\nm}), film “bulk” layer ($t$), and surface roughness layer ($\sigma_\text{RMS}$) modeled by the Bruggemman effective medium approximation using fixed 50\% void volume fraction of voids and film material.  The films were transparent and thus the extinction coefficient is zero in the spectral range used. The refractive index of the films is described by a 2-term Sellmeier formula,
\begin{equation}
n = \sqrt{1+A_\text{UV}\frac{\lambda^2}{\lambda^2-\lambda_0^2}-A_\text{IR}\lambda^2}
\label{eq:sellmeyer}
\end{equation}
where $A_\text{IR}$ is the amplitude of an undamped Lorenz oscillator (pole) with zero resonance energy, $A_\text{UV}$ and $\lambda_0$ are the amplitude and wavelength position of an undamped Lorenz oscillator (pole) in the UV region, respectively.

\begin{table*}[htbp!]
    \def\arraystretch{1.1}
	\caption{Ellipsometry characterization.}
	\label{tab:sellmeyer_fitting}
	\begin{ruledtabular}
    \begin{tabular}{cccrccc}
    Sample & $t$ (\si{\nm}) & $\sigma_\text{RMS}$ (\si{\nm}) & $A_\text{UV}$ & $\lambda_0$ (\si{\um}) & $A_\text{IR}$ (\si{\um}$^{-2}$) \\ \hline
    1     & 48.06 & 2.27 & 1.63816 & 0.1151395 & 0.0146499 \\
    2     & 72.87 & 1.76 & 1.65264 & 0.1122417 & 0.0208134 \\
    3     & 87.91 & 2.93 & 1.67037 & 0.1090871 & 0.0261835 \\
    4     & 120.28 & 2.89 & 1.70322 & 0.0949958 & 0.0369478 \\
    \end{tabular}
    \end{ruledtabular}
\end{table*}

In Figure 1 of the main text, we assumed alumina refractive index to be given by \cref{eq:sellmeyer} with parameters corresponding sample 4 in \cref{tab:sellmeyer_fitting}. The 

Expected $Q$-factor values for uncoated and alumina coated microspheres were performed considering the scattering model of an homogeneous sphere \cite{Vernooy1998, Lin2018},
\begin{equation}
    Q_\text{ss} \simeq \frac{3n^2(n^2+2)^2}{(4\pi)^3(n^2-1)^{5/2}}\frac{\lambda^{7/2}D^{1/2}}{\sigma^2B^2}
\end{equation}
where $n$ is the refractive index of the host material, $\lambda$ is the wavelength, $D$ is the microsphere diameter, $\sigma$ and $B$ are the RMS roughness and its correlation length respectively. \cref{tab:scattering-model} shows scattering $Q$-factors for an uncoated and coated microsphere with $D=\SI{250}{\um}$. The refractive index for alumina was found with \cref{eq:sellmeyer} using parameters for sample 2 in \cref{tab:sellmeyer_fitting}.

\begin{table*}[htbp!]
    \def\arraystretch{1.1}
	\caption{Estimated $Q$-factors using the scattering  model.}
	\label{tab:scattering-model}
	\begin{ruledtabular}
    \begin{tabular}{ccccccc}
    Sphere & $\sigma_\text{RMS}$ (\si{\nm}) & $B$ (\si{\nm}) & $n (\lambda=\SI{1,55}{\um})$ & $Q$-factor \\ \hline
    Uncoated & 0.4 & 90.6 & 1.44 & \num{3e9} \\
    \chemfig{Al_2O_3}-Coated & 4 & 70.7 & 1.62 & \num{2e7} \\
    \end{tabular}
    \end{ruledtabular}
\end{table*}

Surface water formation may be a significant limiting factor. Considering the equation given by Vernooy et al. \cite{Vernooy1998}

\begin{equation}
    Q_\text{water} = \sqrt{\frac{\pi D}{8n^3\lambda}}\frac{1}{\delta\beta_\omega(\lambda)}
\end{equation}
where $\delta$ is the water layer width and $\beta_\omega (\lambda)$ is the absorption coefficient, we can estimate a $Q$-factor of \num{2e7}, taking the same $\delta\approx\SI{0.2}{\nm}$ (1 or 2 monolayers) as used in the reference, $D=\SI{250}{\um}$, and $\beta_\omega(\lambda=\SI{1,55}{\um})=\SI{1181,9}{\m^{-1}}$.

\subsection{Mach-Zehnder interferometer calibration}

To build a reliable frequency axis was necessary to know the dispersion properties of the fiber-MZI. To this end, we characterized the dispersion of the optical fiber (Corning, SMF28) using the chromatic dispersion analyzer (EXFO, FTB 5800). This analyzer, based on the phase-shift method, measure different phase delays in the frequency domain ($\Delta T_g$). The data obtained by this equipment was fitted using the 3-Term Sellmeier equation \cite{Mechels1997AccurateFibers},
\begin{equation}
\label{eq:3term}
    \Delta T_g (\lambda) = \alpha + \zeta\lambda^2 + \frac{\gamma}{\lambda^2}
\end{equation}
where $\alpha, \gamma, \zeta$ are the parameters to be estimated.

\begin{figure*}[htbp!]
\includegraphics{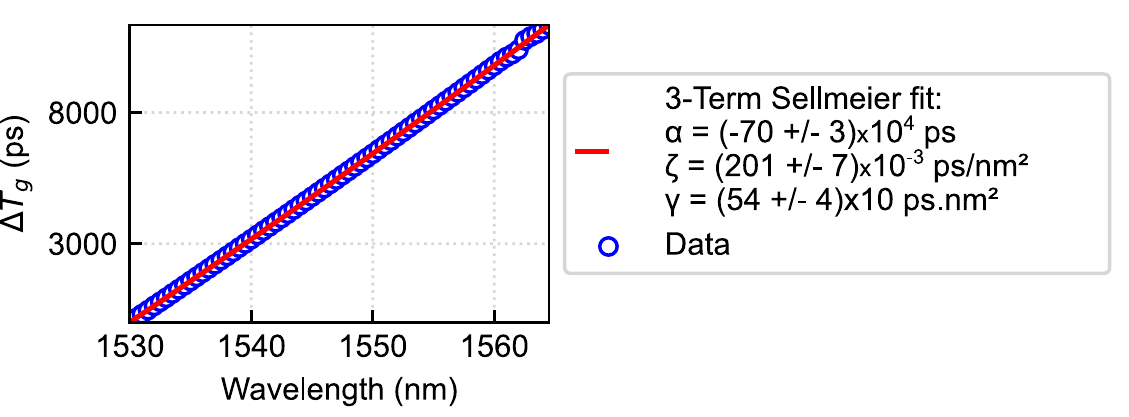}
\caption{Fitting the 3 term Sellmeier equation.}
\label{fig:3-Term-Sellmeier}
\end{figure*}

The fiber group velocity dispersion parameter $D$ can be expressed in terms of the phase delay of a pulse propagating into an optical fiber of length $L$,
 
\begin{equation}
\label{eq:dispersion_3term}
    D = \frac{1}{L}\frac{d\Delta T_g}{d\lambda} = \frac{2\zeta}{L}\left(\lambda - \frac{\gamma}{\zeta}\frac{1}{\lambda^3}\right)
\end{equation}

On the other hand, the group velocity dispersion parameter $D$ in a typical single mode silica fiber is expressed by the empirical relation \cite{SMF-28Corning}:
\begin{equation}
\label{eq:empirical}
    D = \frac{S_0}{4}\left(\lambda-\frac{\lambda_0^4}{\lambda^3}\right)
\end{equation}
where $S_0$ is the zero dispersion slope and $\lambda_0$ is the zero dispersion wavelength.

Comparing equations (\ref{eq:dispersion_3term}) and (\ref{eq:empirical}),
\begin{equation}
\label{eq:S0-Lambda0}
    S_0 = \frac{8\zeta}{L} \hspace{1cm}\text{and} \hspace{1cm} \lambda_0 = \sqrt[4]{\frac{\gamma}{\zeta}}
\end{equation}

Using \cref{eq:S0-Lambda0} and the estimated parameters in \cref{fig:3-Term-Sellmeier} we found $S_0=\SI{82\pm3e-3}{\ps/\km\times\nm^2}$ and $\lambda_0=\SI{1280\pm30}{\nm}$. With this and knowing that $D = -(2\pi c/\lambda^2)\beta_2$ we have the expression for the group velocity dispersion $\beta_2$,

\begin{equation}
    \beta_2 = -\frac{S_0}{8\pi c}\left[\lambda^3 - \frac{\lambda_0^4}{\lambda}\right]
\end{equation}

\begin{figure*}[htbp!]
\includegraphics{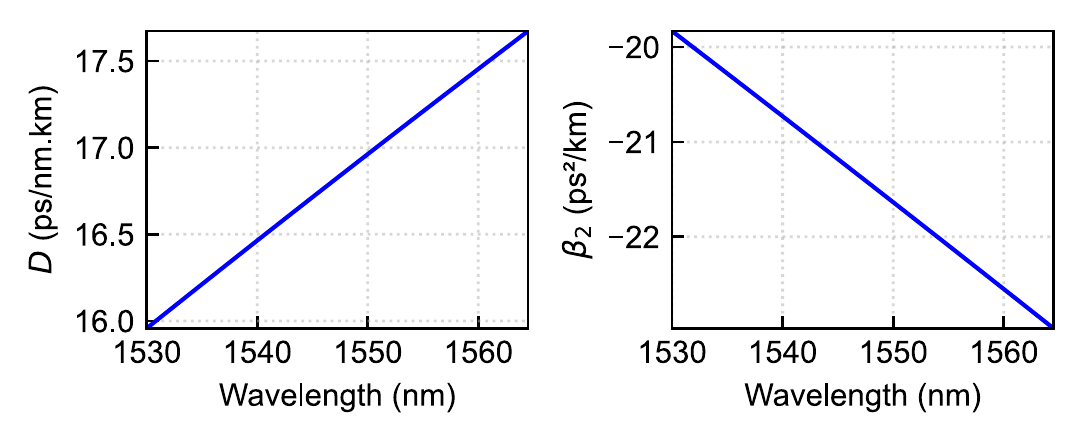}
\caption{Group velocity dispersion parameters $D$ and $\beta_2$ versus wavelength.}
\label{fig:D-and-Beta2}
\end{figure*}

\begin{figure*}[htbp!]
\includegraphics{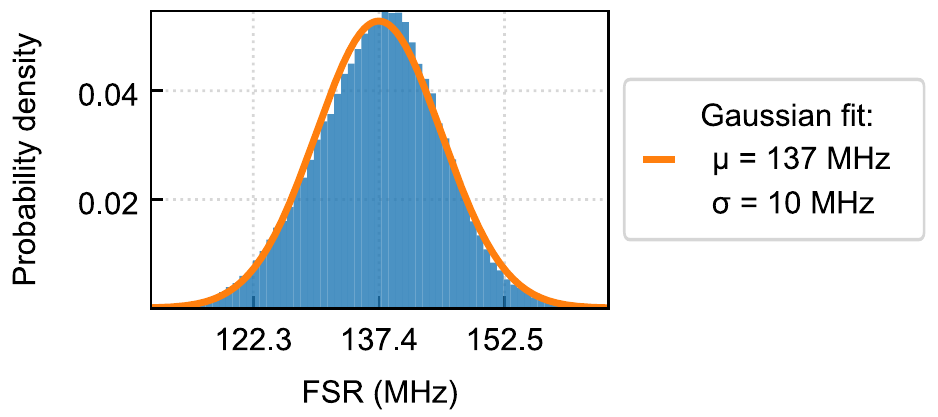}
\caption{Statistic of the FSRs of the fiber-MZI.}
\label{fig:fiber-FSR-histogram}
\end{figure*}

\cref{fig:D-and-Beta2} shows values of $D$ and $\beta_2$ in the telecom wavelengths. Another parameter that can be measure is the equidistant-FSR or $d_1$ of the fiber-MZI. For this, we calibrated the frequency axis using a free-space-MZI that does not suffer dispersion. \cref{fig:fiber-FSR-histogram} shows a statistic of the FSRs of the fiber-MZI where was considered a Gaussian distribution and could thus be found $d_1 = \SI{137\pm10}{\MHz}$.

Substituting the values of $\beta_1 = n_g/c$ , $\beta_2$ and $d_1$ at $\lambda=\SI{1535}{\nm}$ in
$d_2 = -(2\pi\beta_2/\beta_1)d_1^2$ is possible to know the second order dispersion parameter and its uncertainty $d_2 = \SI{0.5\pm0.1}{\Hz}$.

\subsection{\label{section:experiment}Comparing Simulation and Experimental results}

\begin{figure*}[htbp!]
\includegraphics{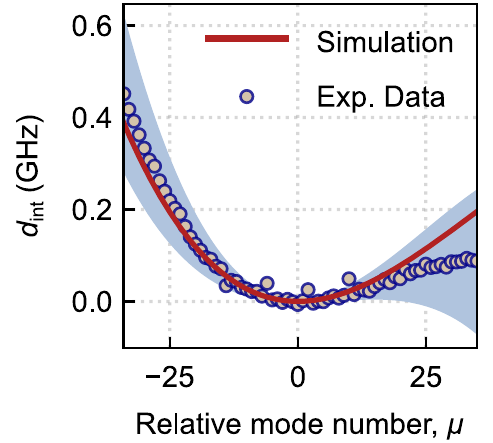}
\caption{Confidence interval.}
\label{fig:confidence}
\end{figure*}

\cref{fig:confidence} shows in blue the confidence interval of the residual dispersion ($d_\text{int}$) for a bare silica microsphere correspond to all possible values that parameters $S_0$ and $\lambda_0$ could take. Experimental data was fitted using the best fitted values of each parameter: $S_0=\SI{82e-3}{\ps/\km\times\nm^2}$ and $\lambda_0=\SI{1280}{\nm}$. The red line here is the simulated result for residual dispersion.

\begin{figure*}[htbp!]
\includegraphics{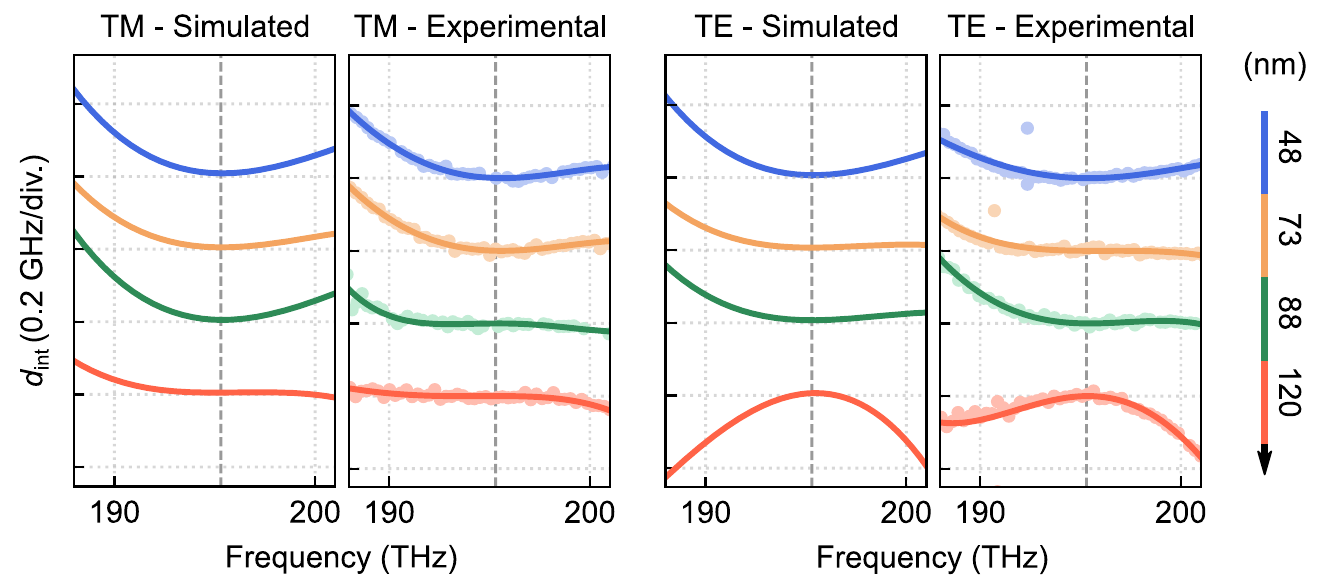}
\caption{Simulated and experimental dispersion curves.}
\label{fig:sim_vs_disp}
\end{figure*}

So, we can build a more reliable frequency axis using the fiber-MZI, via $S_0$ and $\lambda_0$ parameters. With this, we measured the residual dispersion of a series of silica microspheres coated with different thicknesses of the alumina layer. \cref{fig:sim_vs_disp} shows all these curves and makes a comparison of the simulated and experimental fitted curves. Dispersion parameter values ($d_i$) fitted to data around \SI{193.5}{\THz} (\SI{1535}{\nm}) for all the samples are tabulated in \cref{tab:ResumingDispersion}.

\begin{table*}[htbp!]
	\def\arraystretch{1.1}
	\caption{Experimental values for different alumina coating microspheres.}
	\label{tab:ResumingDispersion}
	\begin{ruledtabular}
	\begin{tabular}{ccrrrrrr}
    	& & \multicolumn{3}{c}{TM polarization} & \multicolumn{3}{c}{TE polarization} \\
    	\cline{3-5} \cline{6-8}
		Sample & $D$ (\si{\um}) & $d_2$ (\si{kHz}) & $d_3$ (\si{\kHz}) & $Q_\text{L}$ (\si{\mega}) & $d_2$ (\si{\kHz}) & $d_3$ (\si{\kHz}) & $Q_\text{L}$ (\si{\mega}) \\ \hline
		\def\arraystretch{1}
		0 \footnote{bare silica microsphere.} & \num{248} & \num{488\pm6} & \num{-32\pm1} & \num{13\pm5} & \num{454\pm3} & \num{-26.4\pm0.5} & \num{11\pm3} \\
		1 & \num{238} & \num{297\pm9} & \num{-18\pm1} & \num{10\pm4} & \num{210\pm20} & \num{-9\pm4} & \num{7\pm4} \\
		2 & \num{224} & \num{297\pm8} & \num{-24\pm1} & \num{9\pm3} & \num{80\pm10} & \num{-20\pm2} & \num{9\pm3} \\
		3 & \num{260} & \num{-22\pm40} & \num{-32\pm4} & \num{3\pm2} & \num{180\pm20} & \num{-21\pm3} & \num{3\pm1} \\
		4 & \num{230} & \num{-108\pm6} & \num{-18\pm1} & \num{6\pm3} & \num{-510\pm30} & \num{-32\pm4} & \num{4\pm1} \\
	\end{tabular}
	\end{ruledtabular}
\end{table*}

\bibliography{references_mendeley.bib}